\documentclass[summary]{URSIGASS2021}

\DeclareMathAlphabet{\mathsfbf}{OT1}{cmss}{sbc}{n}

\usepackage{graphicx,amsfonts}
\usepackage[lined,ruled,linesnumbered,noend]{algorithm2e}

\newcommand{\yv}{{\bf y}}

\newcommand{\Hm}{{\bf H}}
\newcommand{\Id}{{\bf I}}

\newcommand{\Mm}{{\bf M}}

\newcommand{\Ym}{{\bf Y}}

\newcommand{\phiv}{\boldsymbol{\phi}}

\newcommand{\Phim}{\boldsymbol{\Phi}}

\def\Herm{\mathsf{^H}}

\def\ben{\begin{enumerate}}
\def\beq{\begin{equation}}
\def\beqa{\begin{eqnarray}}
\def\bit{\begin{itemize}}
\def\een{\end{enumerate}}
\def\eeq{\end{equation}}
\def\eeqa{\end{eqnarray}}
\def\eit{\end{itemize}}

\title{Information aging in massive MIMO systems affected by phase noise}

\author{ Alberto Tarable\affref{ref1},
  and Francisco J. Escribano\affref{ref2}}

\affiliation{%
  \aff{ref1}{CNR-IEIIT, Torino, Italy}
  \aff{ref2}{Universidad de Alcal\'a, Alcal\'a de Henares, Spain}
}

\begin{document}

\maketitle

\begin{abstract}
In massive MIMO systems, phase noise can spoil the performance of the usual receiver techniques. The problem arises because of the aging of phase-noise information based on pilots. In this paper, in a realistic 5G uplink scenario, we quantify the impact of information aging and we propose an iterative receiver based on expectation-maximization (EM). Simulation results show that the iterative receiver is robust to information aging related to phase noise.
\end{abstract}

\section{Introduction}

Massive multi-input multi-output (MIMO) is a key technology to support high data rates in 5G cellular systems. In massive MIMO systems, the base station (BS) is equipped with many antennas, allowing to communicate to several multiantenna users sharing the same time-frequency resources. With ideal channel state information (CSI), in the uplink, the symbols coming from all users can be correctly demodulated by properly filtering the received signal, in order to reduce the interference among different users. 

The crucial problem for massive MIMO systems is thus acquiring a sufficiently good CSI, which is a nontrivial task, due to the large number of antennas involved both at the transmitter and at the receiver side. Typically, pilots are periodically transmitted to estimate the channel, and the latest available channel estimation is used to filter the received signal and demodulate the user symbols. Consequently, if the channel changes (due to, e.g., mobility) between two successive pilot transmissions, data are demodulated   
based on obsolete CSI, i.e., \emph{information aging} takes place.

This is especially true whenever phase noise at the transmitter and at the receiver side affects the signal.
Phase noise typically has a faster dynamics than channel fading, and its impact on the performance of filtering at the BS may impair correct data demodulation. It becomes thus of paramount importance to evaluate the impact of CSI aging due to phase noise. This is the subject of several works from the literature, in different scenarios (\cite{Krishnan}-\cite{Corvaja}). Hovewer, most of such works do not propose phase-detection techniques to mitigate the effect of phase noise.

In this work, we investigate the application of the iterative phase detector described in~\cite{Tarable} to the uplink of massive MIMO systems. Since the phase detector is applied at the BS receiver, we assume that it is able to cope with the additional computational burden. By simulation, we show the benefit of phase recovery and the impact of the model parameters to the overall performance.  

The remaining of the paper is organized as follows. In Section 2, we describe the massive MIMO system and the frame structure. In Section 3, we introduce the EM-based iterative receiver. In Section 4, we show simulation results to assess the performance of the proposed receiver. Finally, in Section 5, we draw some conclusions. 

\section{System description}

Consider an $N_t \times N_r$ massive MIMO channel with $O_t$ oscillators at the transmitter and $O_r$ oscillators at the receiver. Each transmit-side oscillator feeds $N_{o,t} = N_t / O_t$ antennas, while  $N_{o,r} = N_r / O_r$ receive antennas are fed by the same oscillator. Different oscillators introduce independent phase noise. The input-output relationship at time $n=1,2,\,\dots$ is given by
\begin{equation} \label{eq:cha}
\mathbf{y}[n] = \Phim_R[n]\, \mathbf{H}[n]\, \Phim_T[n] \mathbf{x}[n] +
\mathbf{z}[n]
\end{equation}
where:
\begin{itemize}
\item $\mathbf{H}[n]$ is the $N_r \times N_t$ channel matrix at time $n$;

\item $\Phim_T[n] = \mathrm{diag}\left(e^{j\phi_1[n]}, \dots,  e^{j\phi_{O_t}[n]}\right) \otimes \Id_{N_{o,t}}$ and $\Phim_R[n] = \mathrm{diag}\left(e^{j\phi_{O_t+1}[n]}, \dots,  e^{j\phi_{O_t+O_r}[n]}\right)  \otimes \Id_{N_{o,r}}$ are the diagonal matrices of transmit and receive phase-noise coefficients at time $n$, respectively, assumed to be unknown at both sides; 

\item $\mathbf{x}[n]$ and $\mathbf{y}[n]$ are the column vectors of the $N_t$ transmitted symbols and $N_r$ received samples at time $n$, respectively;

\item $\mathbf{z}[n]$ is a size-$N_r$ vector of zero-mean, circularly-invariant Gaussian-noise samples, with variance $\sigma^2$ per real dimension, which are supposed to be independent across time and receive antenna.
\end{itemize}

For the phase-noise samples, time dependency is kept into account by assuming Wiener phase-noise processes:
\begin{equation} \label{eq:wiener}
\phi_i[n] = \phi_i[n-1] + w_i[n], \,\,\,i=1,\dots,O_r+O_t,\,n=1,2,\dots
%\mathrm{mod}\,\, 2\pi,
\end{equation}
where $\phi_1[0],\dots,\phi_{O_r+O_t}[0]$ are independent and uniformly distributed over $[0,2\pi)$ and $w_i[n],\dots,w_{O_r+O_t}[n]$, are independent zero-mean white Gaussian processes with power $\rho^2$ (all processes have the same power).

Let us notice that each tap of the MIMO channel is affected by a \emph{sum} phase-noise process:
\begin{equation}
\phi_{ii'}[n]  = 
{\phi}_{i}[n] +
{\phi}_{O_t+i'}[n] ,\,\,\, i=1,\dots,O_t,\,i'=1,\dots,O_r 
\end{equation} 
which is the sum of one transmit and one receive {\em atomic} phase-noise process.
We define for future use the size-$(O_t+O_r)$ vector $\phiv[n]$, whose $i$-th element is $\phi_i[n]$, $i=1,\dots, O_r+O_t$. 
%Analogously, we define the size-$(O_tO_r)$ vector $\phiv^{\mathrm{sum}}[n]$, whose element $(i-1)O_r + i'$ is $\phi_{ii'}[n]$, $i=1,\dots, O_t$, $i'=1,\dots, O_r$. 

\begin{figure}
\includegraphics[width=1.0\columnwidth]{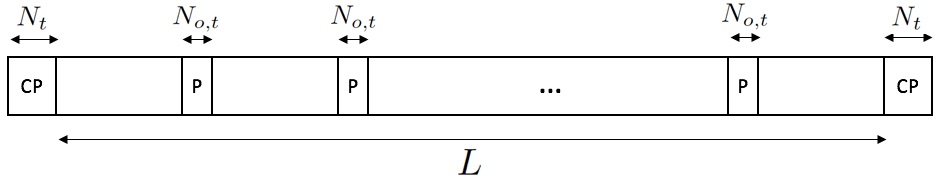}
\caption{Frame structure.}\label{fig:figure_1}
\end{figure}

\subsection{Pilot-based channel estimation}

%The underlying hypothesis behind the channel model in \eqref{eq:cha} is that the channel matrix $\Hm[n]$ changes in time much more slowly than the phase noise. Thus, channel estimation can be performed less frequently than phase detection. In particular, we will suppose that there are two kinds of pilots in the system, with the first pilot type devoted to joint channel and phase estimation (as it is not possible, in principle to distinguish the phase of unknown channel matrix entries from phase noise), and the second devoted only to phase detection.

Fig.~\ref{fig:figure_1} shows the structure of the transmitted frame. Every $L$ channel uses, channel estimation is performed through the transmission of $N_t$ channel pilots (labelled ``CP'' in the figure).
The channel matrix $\widehat{\Hm}$ estimated at the beginning of a data frame is used for the whole frame, although the channel may experience variation between two channel estimation stages. 

Every $R$ data channel uses, other pilots are transmitted (labelled ``P'' in the figure), which are used for phase detection only, since phase noise is assumed to vary much faster than channel coefficients.  
Such pilots take $O_{t}$ channel uses, during which only one group of transmit antennas fed by the same oscillator are switched on, while the others are switched off.

The parameter $R$ must be chosen according to a trade-off between two conflicting exigences. $R$ should be chosen as large as possible in order to increase the transmission efficiency, defined as
\beq
\eta = \frac{R}{R + O_{t}}
\eeq 
On the other side, as shown in Figures \ref{fig:figure_3}-\ref{fig:figure_4}, phase-noise information aging takes place between two successive phase estimation stages, and the larger $R$, the more relevant is the impact of this information aging on receiver performance. So, $R$ should be kept not too large in order to avoid exceeding a prescribed value of the \emph{Age of Information} (AoI), defined in this paper as
\beq
\mathrm{AoI} = \frac{R \rho^2}{1 \mbox{ degree}^2}
\eeq
The above definition, which is equal to the variance of the Wiener increment of phase noise between two successive pilot transmission stages, is proposed to take into account both the inter-pilot data burst length and the phase noise dynamics.  

\section{EM-based receiver}

%In the following, we will denote with $\Hm_{ij}$, $i = 1,\dots,O_r$, $j = 1,\dots,O_t$,  the  $ N_{o,r} \times N_{o,t}$ block of $\Hm$ corresponding to oscillators $j$ and $N_t + i$ at the transmitter and receiver side, respectively. Moreover, $\Hm_{i}$, $i=1,\dots,O_t$ will be the $N_r \times N_{o,t}$ matrix obtained by stacking all matrices $\Hm_{ji}$ for $j = 1,\dots, O_r$. Finally, we will denote with  $\Hm_{O_t + i}$, $i=1,\dots,O_r$ the $N_{o,r} \times N_t$ matrix obtained by juxtaposing all matrices $\Hm_{ij}$ for $j = 1,\dots, O_t$. With $\widehat{\Hm}_{ij}$, $\widehat{\Hm}_{i}$ and $\widehat{\Hm}_{O_t + i}$ we will denote the corresponding estimated submatrices.

\begin{figure}
\includegraphics[width=1.0\columnwidth]{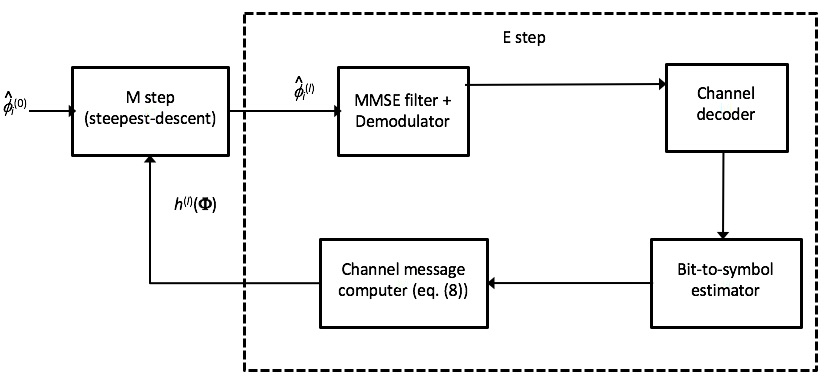}
\caption{Block scheme of the proposed EM receiver.}\label{fig:figure_2}
\end{figure}

Suppose a frame of $L$ symbol vectors $\mathbf{X}=(\mathbf{x}[1],\dots,\mathbf{x}[L])$ is transmitted through the channel described by (\ref{eq:cha}) and let $\Ym = (\yv[1],\dots,\yv[L])$ be the channel output. Also, define $f_A(\Phim)$  the  a priori distribution of $\Phim = (\phiv[1],\dots,\phiv[L])$. As in~\cite{Tarable}, we consider an EM-based iterative receiver, whose block scheme is shown in Fig. \ref{fig:figure_2}.

At iteration $l$, $l=1,2,\dots$, the EM algorithm performs the following two steps:
\begin{itemize}
\item[\bf{E step:}] The average over transmitted symbols is computed as follows:
\begin{equation} \label{eq:Estep}
h^{(l)}(\Phim)= E_{\mathbf{X}}^{(l)} \log \prod_{n=1}^{L}\Pr\left\{ \mathbf{y}[n] |  \mathbf{x}[n] , \mathbf{\phi}[n]\right\}
\end{equation}
where the average is performed according to the distribution $\Pr \{ \mathbf{X}| \mathbf{Y},\widehat{\Phim}^{(l-1)}\}$ for $l > 1$ and to the a-priori symbol distribution $\Pr \{ \mathbf{X}\}$ at the first iteration. 

\item[\bf{M step:}] The following maximization problem is solved:
\begin{equation} 
\widehat{\Phim}^{(l)} = \arg \max_{\Phim}  \left( h^{(l)}(\Phim) + \log f_A(\Phim)
\right)
\end{equation}

\end{itemize}

For the channel model in (\ref{eq:cha}), apart from an inessential additive constant, (\ref{eq:Estep}) becomes:
\begin{equation} \label{eq:Estep1}
h^{(l)}(\Phim)= \sum_{n=1}^{L} \frac{\Re\{\widetilde{\mathbf{x}}^H[n] \Phim_T^H[n] 
\widehat{\mathbf{H}}^H \Phim_R^H[n] \mathbf{y}[n]\}}{\sigma^2}
\end{equation}
where $\widetilde{\mathbf{x}}[n] = E_{\mathbf{X}}^{(l)} \mathbf{x}[n]$. 

The a-posteriori probability on input symbols $\Pr \{ \mathbf{X}| \mathbf{Y},\widehat{\Phim}^{(l-1)}\}$ is approximated through standard demodulation and possibly (if the transmitted message is encoded) decoding, for the current estimated value of phase noise. More precisely, we suppose that demodulation is performed by first applying a linear minimum mean-square error (LMMSE) filter in order to reduce the multiuser interference, and then by single-user demodulation of the LMMSE output. The LMMSE filter matrix at the $l$-th iteration is given by:
\beq
\Mm^{(l)} = \left( \widetilde{\Hm}[n]\Herm \widetilde{\Hm}[n] + \sigma^2 \Id \right)^{-1} \widetilde{\Hm}[n]\Herm
\eeq 
where
\beq
\widetilde{\Hm}[n] = \widehat{\Phim}_R^{(l)}[n] \widehat{\Hm} \widehat{\Phim}_T^{(l)}[n]
\eeq

As in \cite{DauKoLoe05a}, the maximization involved in the M step is approximated through the steepest-descent algorithm. Let $\widehat{\Phim}^{(l)}_m$ be the estimate of $\widehat{\Phim}^{(l)}$ after $m$ steepest-descent iterations. Let the starting point be $\widehat{\Phim}^{(l)}_0 = \widehat{\Phim}^{(l-1)}$ for $l > 1$ and $\widehat{\Phim}^{(1)}_0 = \widehat{\Phim}_P$, where $\widehat{\Phim}_P$ is an initial estimate based on pilots (see \cite{Tarable} for more details). Then:
\begin{equation}
\widehat{\Phim}^{(l)}_{m+1} = \widehat{\Phim}^{(l)}_{m} + \lambda \nabla_{\Phim}
\left( h^{(l)}(\Phim) + \log f_A(\Phim) \right)\Bigg|_{\widehat{\Phim}^{(l)}_m}
\end{equation}
where $\lambda$ is the step size of the steepest-descent algorithm and can be optimized numerically.
See \cite{Tarable} for the explicit computation of the above gradient. 
%The partial derivative with respect to $\phi_i[n]$, $i=1,\dots,O_t$ of the first term in parentheses is readily computed as:
%\begin{equation} \label{eq:deri1}
%\frac{\partial}{\partial \phi_i[n]} h^{(l)}(\Phim) = \frac{\Im\{e^{-j \phi_i[n]}\widehat{\xv}_i^H[n]  \widehat{\mathbf{H}}_i^H \Phim_R^H[n] \mathbf{y}[n]\}}{\sigma^2}
%\end{equation}
%while an analogous computation holds for the derivative with respect to $\phi_{O_t+i}[n]$, $i=1,\dots,O_r$.
%For the derivative of the second term in parentheses, thanks to the Wiener model, we have with a slight abuse of notation:
%\begin{equation}
%\log f_A(\Phim) = \log f_A(\mathbf{\phi}[0]) + \sum_{n=1}^{L} \log f_A(\mathbf{\phi}[n+1] | \mathbf{\phi}[n])
%\end{equation}
%so that, provided that $\rho^2 \ll 2\pi$ (a usually realistic approximation), we obtain (\cite{DauKoLoe05a}):
%\begin{eqnarray}
%\frac{\partial}{\partial \phi_i[n]} \log f_A(\Phim) &=& \frac{\partial}{\partial \phi_i[n]} \log f_A(\Phim[n] | \Phim[n-1])+ \nonumber\\
%&& \frac{\partial}{\partial \phi_i[n]} \log f_A(\Phim[n+1] | \Phim[n]) \nonumber \\
%& \simeq & \frac{\phi_i[n-1]-\phi_i[n]+k_{n-1}2\pi}{\rho^2}+ \nonumber\\
%&&  \frac{\phi_i[n+1]-\phi_i[n]+k_{n}2\pi}{\rho^2}
%\end{eqnarray}
%where $k_n$ and $k_{n-1}$ are signed integers that minimize the moduli of the numerators.

\section{Simulation results}

In this section, we show simulation results for the system described in the previous sections and higlight the benefit in terms of information aging due to the adoption of the EM-based receiver.

We consider a system with $K = 16$ users, each equipped with a two-antenna user equipment. Thus, we have $O_t = K = 16$ independent oscillators, each feeding $N_{ot} = 2$ antennas, for a total of $N_t = 32$ transmit antennas. The BS is equipped with $O_r = 8$ oscillators, each one feeding $N_{or} = 8$ antennas, for a total of $N_r = 64$ receive antennas.

Each user encodes its information bit stream with a 5G NR LDPC code of type 2 and lift factor $Z = 2$, with a length $N = 104$ bits and a rate equal to $0.8$ \cite{5G}. Each frame transmits 1000 LDPC codewords. The employed modulation format is 256-QAM. The transmit power is normalized to 1 so that the signal-to-noise ratio on the channel is $E_s/N_0 = (2\sigma^2)^{-1}$. 

 In the simulations, we will suppose that the channel matrix is fixed for the whole frame and that channel estimation is perfect, i.e. $\widehat{\Hm} = \Hm$. Regarding phase-noise processes, the standard deviation of the increment is equal to $\rho = \sqrt{0.2}$ degrees. As a consequence of channel estimation, we can safely set the phase-noise values to zero at the beginning of the frame, i.e., $\widehat{\phi}_i[0] = 0$, $i = 1,\dots, O_r+O_t$. 

At the receiver, the LDPC decoder implements the optimal BP algorithm with at most 50 iterations and genie-aided stopping rule. The phase detector performs at each receiver iteration 5 steps of the steepest-descent algorithm with a step size $\lambda = 2.5 \times 10^{-4}$. The computation of \eqref{eq:Estep1} is slightly simplified by substituting $\widetilde{\mathbf{x}}[n]$ with the hard estimates  $\widehat{\mathbf{x}}[n]$ of the transmitted symbols. This simplification has been shown in \cite{Tarable} to have a negligible effect on performance. The decoder performs at most 10 iterations of the EM algorithm, but it stops earlier if the LDPC decoder triggers its stopping rule. For each $E_s/N_0$ value, we count 100 frame errors.

\begin{figure}
\includegraphics[width=1.0\columnwidth]{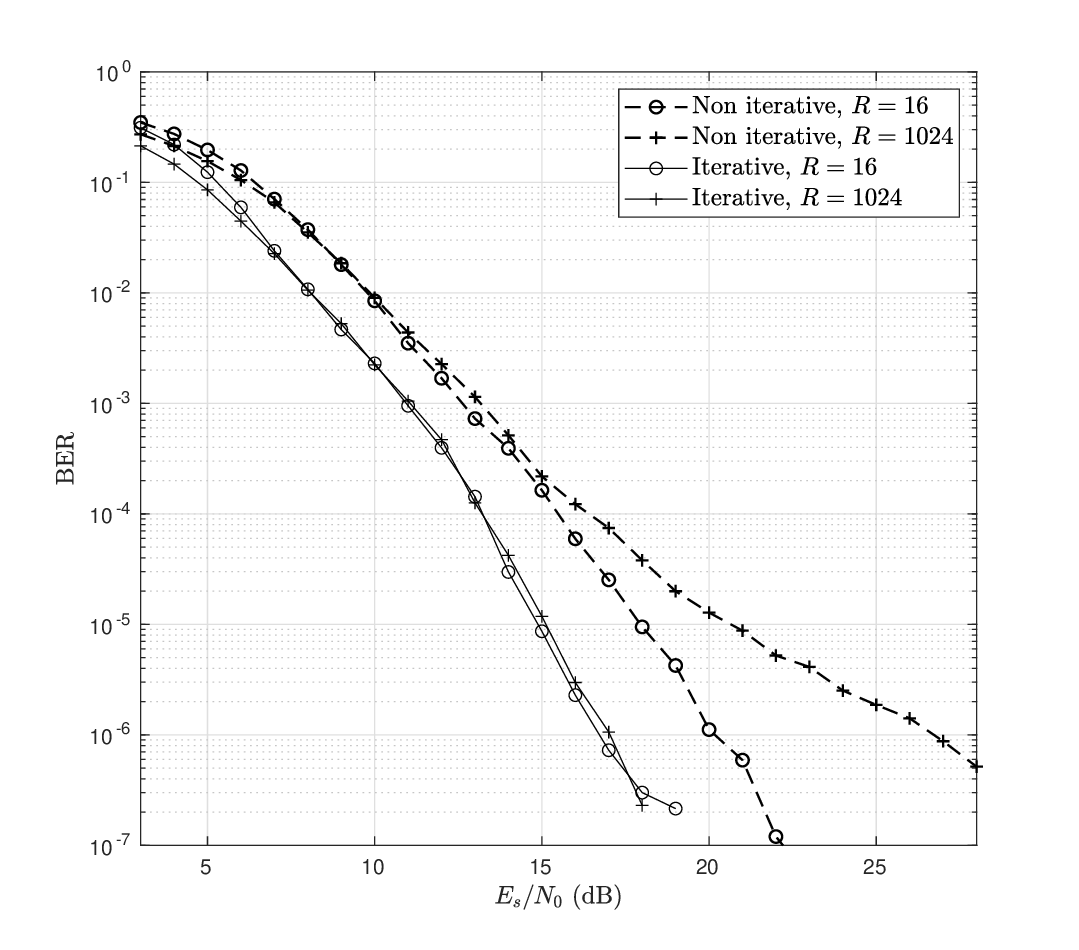}
\caption{Performance of the proposed EM receiver with different values of $R$. AWGN channel.}\label{fig:figure_3}
\end{figure}

In Fig. \ref{fig:figure_3}, we show the performance of the receiver on the AWGN channel (solid lines), for different values of $R$, the length of the inter-pilot data burst. All entries of the channel matrix have magnitude 1. The line with circles is for $R = 16$, which corresponds to a low efficiency $\eta = 50 \%$ and a low $\mathrm{AoI} = 3.2$. The line with crosses is for $R = 1024$, corresponding to a high efficiency $\eta = 98.5 \%$ and a high $\mathrm{AoI} = 204.8$. The increase in AoI is well tolerated by the system as the performance degradation in passing from $R = 16$ to $R = 1024$ is less than 1 dB. As a comparison, in the figure, we have also plotted (dashed lines) the performance of a scheme in which we simply estimate the phase noise on the basis of pilots, without any feedback from the decoder to the phase detector. The results show that, in this case, information aging is apparent not only in the $E_s/N_0$ loss, but also in a lower slope of the $R= 1024$ curve with respect to the case $R = 16$. 

\begin{figure}
\includegraphics[width=1.0\columnwidth]{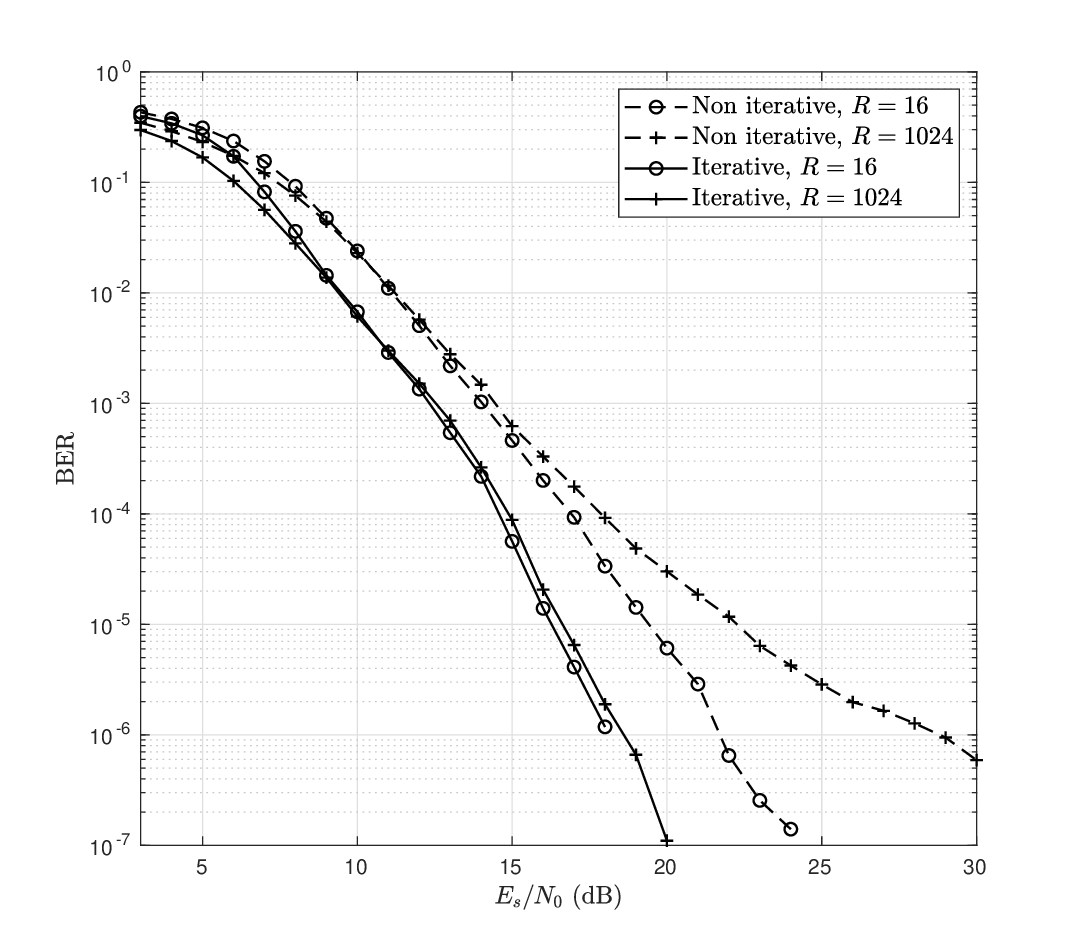}
\caption{Performance of the proposed EM receiver with different values of $R$. Rice-fading channel with $K = 0$ dB.}\label{fig:figure_4}
\end{figure}

In Fig. \ref{fig:figure_4}, we show the performance of the receiver on a Rice-fading channel with  a Rice factor $K = 0$ dB. The channel matrix is constant for the whole frame. We can draw the same conclusions as for the AWGN channel. The dashed curves, which correspond to the non iterative receiver, show the dramatic effect of information aging in going from $R = 16$ to $R = 1024$. The iterative receiver is much more robust to information aging, with the more efficient $R = 1024$ case losing only a fraction of dB with respect to the other. Comparing  Fig. \ref{fig:figure_4} with Fig. \ref{fig:figure_3}, we can see that the iterative receiver loses only a couple of dB due to fading. 

% \begin{figure}
% \includegraphics[width=1.0\columnwidth]{BCRB_comparative}
% %\includegraphics[width=0.495\textwidth]{figures_1/Link1_HOP_M4_A1_1.eps}
% \caption{Mean square error (MSE) of the proposed EM receiver with different values of $R$, both under the AWGN channel and the Rice-fading channel.}\label{fig:figure_5}
% \end{figure}

%Finally, in Fig. \ref{fig:figure_5}, we show the mean square error (MSE) of the phase-noise estimates in the different simulated scenarios. We average MSE over all processes, samples and channel realizations. We show only the average MSE obtained by the iterative receiver. As it can be seen, the performance is very similar for the AWGN channel (solid lines) and the Rice-fading channel (dashed lines) with the same value of $R$. Going from $R = 16$ to $R = 1024$ entails a worsening of the MSE performance of about a factor of 5. As a benchmark, we also show the Bayesian Cram\'er-Rao bound (BCRB), whose expression can be obtained by an extension of the computation in \cite{Tarable2}. As it can be seen, the simulated curves show the same trend as the BCRB, with the $R = 1024$ case losing about an order of magnitude with respect to the benchmark.  

\section{Conclusions}

In this paper, we have quantified the impact of information aging in a  realistic massive-MIMO 5G uplink scenario. We have also proposed an iterative receiver based on expectation-maximization (EM), robust to information aging related to phase noise.

Future work can consider the impact of imperfect channel estimation on the system performance, the optimization of the receiver parameters, the performance analysis and the scaling law of performance with the number of antennas.


\begin{thebibliography}{99}

\bibitem{Krishnan} R. Krishnan \textit{et al.}, ``Linear Massive MIMO Precoders in the Presence of Phase Noise -- A Large-Scale Analysis,'' \textit{IEEE Trans. on Veh. Tech.}, vol. 65, no. 5, pp. 3057-3071, May 2016.
\bibitem{Pitarokoilis} A. Pitarokoilis, S. K. Mohammed and E. G. Larsson, ``Uplink Performance of Time-Reversal MRC in Massive MIMO Systems Subject to Phase Noise,'' \textit{IEEE Trans. on Wirel. Comm.}, vol. 14, no. 2, pp. 711-723, Feb. 2015.
\bibitem{Corvaja} R. Corvaja and A. G. Armada, ``Phase Noise Degradation in Massive MIMO Downlink With Zero-Forcing and Maximum Ratio Transmission Precoding,'' \textit{IEEE Trans. on Wirel. Comm.}, vol. 65, no. 10, pp. 8052-8059, Oct. 2016.
\bibitem{Tarable} A. Tarable, G. Montorsi, S. Benedetto, and S. Chinnici, ``An EM-based phase-noise estimator for MIMO systems,'' in \textit{Proc. IEEE Int. Conf. Commun. (ICC)}, Budapest (Hungary), Jun. 2013.
\bibitem{DauKoLoe05a} J. Dauwels, S. Korl, and H.-A. Loeliger, ``Expectation maximization for phase estimation,'' in \textit{Proc. of the 8th Int. Symp. on Commun. Th. and Appl.},  2005.
\bibitem{5G} 3GPP TS 38.212. ``NR; Multiplexing and channel coding.'' 3rd Generation Partnership Project; Technical Specification Group Radio Access Network.
\bibitem{Tarable2} A. Tarable, C. Camarda and G. Montorsi, ``Cramer-Rao bounds for MIMO LOS systems affected by distributed Wiener phase noise in the large-blocklength regime'', in \textit{Proc. 2014 IEEE Int. Symp. on Contr., Comm. and Sig. Proc. (ISCCSP)}, May 2014.

\end{thebibliography}
\end{document}